\documentclass[10pt,leqno]{article}
\usepackage{graphicx}
\usepackage{indentfirst,csquotes}

\usepackage{amssymb,amsthm,amsmath}
\usepackage{xcolor,paralist,hyperref,titlesec,fancyhdr,etoolbox}

\usepackage{siunitx}
\usepackage{chemformula}

\hypersetup{ colorlinks=true, linkcolor=black, filecolor=black, urlcolor=black }

\usepackage{lipsum}

\usepackage{authblk}

\begin{document}
\title{Characterization of heat transfer in 3D CMOS structures using Sideband Scanning Thermal Wave Microscopy}

\author[1]{Valentin Fonck}
\author[1]{Mohammadali Razeghi}
\author[1]{Jean Spièce}
\author[2]{Phillip~Dobson}
\author[2]{Jonathan~Weaver}
\author[3]{George Ridgard}
\author[3]{Grayson M. Noah}
\author[1]{Pascal Gehring\thanks{pascal.gehring@uclouvain.be}}
\affil[1]{Institute of condensed matter and nanosciences (IMCN), Université Catholique de Louvain, Louvain-la-Neuve, 1348, Belgium}
\affil[2]{James Watt School of Engineering, University of Glasgow, Glasgow, G12 8LT, United Kingdom}
\affil[3]{Quantum Motion, 9 Sterling Way, London N7 9HJ, United Kingdom}

\maketitle
\fbox{\begin{minipage}{\linewidth}
This work has been submitted to the IEEE for possible publication. 
Copyright may be transferred without notice, after which this version may no longer be accessible.
\end{minipage}}

\begin{abstract}
Efficient thermal management is critical for cryogenic CMOS circuits, where local heating can compromise device performance and qubit coherence. Understanding heat flow at the nanoscale in these multilayer architectures requires localized, high-resolution thermal probing techniques capable of accessing buried structures.

Here, we introduce a sideband thermal wave detection scheme for Scanning Thermal Microscopy, S-STWM, to probe deeply buried heater structures within CMOS dies. By extracting the phase of propagating thermal waves, this method provides spatially resolved insight into heat dissipation pathways through complex multilayer structures. Our approach enables quantitative evaluation of thermal management strategies, informs the design of cryo-CMOS circuits, and establishes a foundation for \textit{in situ} thermal characterization under cryogenic operating conditions.
\end{abstract}

\section{Introduction}
Quantum computers hold the promise of solving computational problems intractable for classical computers. Recent advancements have pushed the number of physical qubits in a single system beyond the one-thousand mark \cite{abughanem_ibm_2025} but have also  highlighted significant challenges in scalability, integration, and thermal management. In particular, connecting each qubit to room-temperature equipment via individual cables would result in thermal loads and space requirements exceeding the capacity of current commercially available dilution refrigerators --- the so-called I/O bottleneck of quantum computing. A promising approach to mitigate this issue is to position certain measurement and control electronics at cryogenic temperatures (e.g. cryo-CMOS) in close proximity to the qubits such that the I/O count of cabling up to higher-temperature stages is greatly reduced \cite{acharya2022overcoming,JvS2025snh,pauka2021cryocmos,intel2024pandotree}.

However, this proximity necessitates the development of efficient thermal management solutions with minimal spatial footprint as operating cryo-CMOS circuits within dilution refrigerators can lead to significantly elevated temperatures relative to their ultra-cold surroundings \cite{de_kruijf_measurement_2024,noah_cmos_2024,t_hart_characterization_2021}. These temperature gradients induced on thermally coupled systems result in reduced coherence times for neighbouring qubits~\cite{huang_qubit_2024}. Addressing this issue will demand the development of advanced packaging and integration strategies along with thermally aware design and layout of cryo-CMOS circuits, potentially even allowing for co-integration of CMOS circuitry on the same chip as the physical qubits. Knowledge of deep-cryogenic material properties, especially in the sub-4 K temperature range, is extremely limited. Even if some bulk properties of relevant materials are known for the temperature range of interest, commercial CMOS foundries often do not fully reveal all materials, thicknesses, and semiconductor doping concentrations used in their process to their customers or Process Design Kit (PDK) users. Therefore, the effective thermal properties of cryo-CMOS integrated circuits must be determined empirically.

To enable this \textit{in situ} cryogenic thermal characterization of CMOS technologies and devices at their base operating temperatures, further development of cryo-capable, localized characterization tools is necessary \cite{Fonck_2025}. While integrated on-chip heater and thermometer devices in the front-end-of-line (FEOL) allow for characterization of local device self-heating and intra-chip inter-device heating \cite{noah_cmos_2024}, the back-end-of-line (BEOL) remains largely uncharacterized for deep-cryogenic conditions and presents an increasingly important aspect as dense packaging and 2.5D/3D integration become required to achieve quantum systems of larger scale. In particular, the thermal pathways between chip surface metal (e.g. pads, bumps) and buried devices must be characterized \cite{woon_thermal_2025}. The thermal behaviors of these pathways are challenging to predict due to a complex multi-layer structure consisting of thermal conductivities with cubic temperature relationships (e.g. Si \cite{thompson1961thermal}, amorphous SiO$_2$ \cite{pohl2002low}, and Si-Cu boundaries \cite{swartz1989thermal}) and linear temperature relationships (e.g. metals such as Cu and Al \cite{ekin_experimental_2007}). Extreme power densities \cite{pop2010energy} and generally much lower thermal conductivities compared to room temperature \cite{thompson1961thermal} act to generate self heating in transistors which can exceed 90~K at a 6.5~K base temperature \cite{artanov2022self} and consequently induce massive on-chip temperature gradients \cite{pop_heat_2006}. Added dimensional effects (e.g. when the silicon dimension approaches the phonon mean free path length \cite{shiomi2024scientific}) compound this challenge, as BEOL structures can span from nanometer- to millimeter-scale. 

Characterizing and modeling these thermal pathways will facilitate the evaluation and optimization of traditional thermal management strategies such as deep thermal through-silicon vias (TSV) \cite{Feng_TSV} and BEOL heat sinks \cite{halder2023beolheatsink} for the cryogenic environment. This will also enable more sophisticated design and development of cryo-specific thermal management approaches, e.g. making use of superconducting interconnects \cite{thomas2022interposer}.

Among the most promising techniques for enabling this characterization is Scanning Thermal Microscopy (SThM) due to its versatility, compatibility with rough surfaces, ability to measure thermal conductivities, and potential for operation at cryogenic temperatures \cite{Fonck_2025,swami_experimental_2024,bourgeois_liquid_2006}. SThM is a variant of contact-mode Atomic Force Microscopy (AFM) in which a temperature-sensing element is integrated at the apex of the AFM probe. A feedback mechanism -- such as laser reflection on a photodiode or phase modulation of a tuning fork -- is used to control the force applied by the cantilever during scanning. As the probe is scanned over the sample surface, minute changes in the resistance of the integrated thermometer provide spatially resolved information on thermal conductance and/or local temperature. This technique enables researchers to investigate nanoscale heating and heat diffusion phenomena. As a result, SThM has been widely employed to study hot spots in active devices \cite{menges_quantitative_2012,kim_quantitative_2011,kim_ultra-high_2012,saidi_scanning_2009,saidi_tuning_2011,shi_thermal_2001}, as well as to analyze device failure mechanisms \cite{fiege_failure_1998,fiege_thermal_1999,gomes_temperature_2007}. However, accurately probing surface heating from deeply buried heater structures in a CMOS chip poses an experimental challenge that extends beyond simply maximizing the sensitivity of the measurement scheme. 

In this work, we address this challenge by introducing a sideband thermal wave detection scheme for SThM (see Figure \ref{fig:method}). This approach builds on previously developed AC-SThM techniques for quantitative surface thermometry, which mitigate the influence of local variations in thermal contact resistance between probe and sample \cite{Menges2016,Harzheim2018}. By detecting sidebands, we gain access to the phase information of the propagating thermal wave. This phase response encodes the out-of-equilibrium thermal dynamics within the sample. Such knowledge is essential for understanding heat dissipation pathways in complex device architectures and for guiding the design of next-generation low-power electronic technologies.

\section{Methodology and Sample Description}
\begin{figure}
    \centering
    \includegraphics[width=0.85\linewidth]{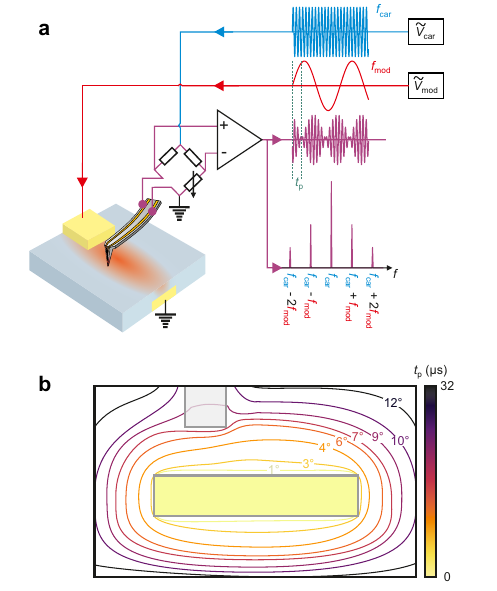}
    \caption{\textbf{a}~Schematic of the sideband-SThM setup used in this work. A high-frequency carrier signal $f_\mathrm{car}$ probes the resistance of the thermometer embedded in a Wheatstone bridge, while a low-frequency excitation $f_\mathrm{mod}$ is applied to a heater buried beneath the sample surface. The resulting SThM signal at the Wheatstone bridge is an amplitude-modulated $f_\mathrm{car}$, which can be demodulated to extract the sideband signals. Input and output signals are shown schematically and are not to scale. \textbf{b}~Isophase curves for the heat waves propagating around a buried heating element. Another material with a different thermal diffusivity is shown in grey. The added colorbar shows the conversion from phase to propagation time of the heat wave. Color scale indicates the conversion to a propagation time.}
    \label{fig:method}
\end{figure}

In our Sideband Scanning Thermal Wave microscopy (S-STWM), an on-sample heating element is modulated at a low frequency $f_\mathrm{mod} = 1$~kHz while the SThM probe is monitored at a higher carrier frequency $f_\mathrm{car} = 91$~kHz. The on-sample heater thus induces a low-frequency variation in the surface temperature, which in turn amplitude-modulates the high-frequency signal detected by the probe, as illustrated in Fig.~\ref{fig:method}a. This mixing process produces sidebands that encode the local temperature oscillations induced by the heating element. The amplitude of the resulting sideband signal is proportional to the local temperature oscillations while the phase is linked to the propagation time of the heat wave from the heating element to the SThM probe's sensor.

In prior work \cite{Harzheim2018}, tandem demodulation was used to obtain amplitude and phase of such temperature oscillations, where the $f_\mathrm{car}$ signal is demodulated at $f_\mathrm{mod}$. While effective, this approach has two drawbacks: (i) if the first-stage bandwidth is narrower than $f_\mathrm{mod}$, the modulation signal may be attenuated or lost, especially given the long integration times typical for thermal signals; (ii) the recovered amplitude depends on the relative phase of the two demodulators, requiring additional post-processing to obtain the true sideband amplitude.

To overcome the limitations of tandem demodulation, we implement a parallel synchronized demodulation scheme. In this approach, the raw signal is simultaneously fed into three phase-locked demodulators operating at $f_\mathrm{car}$, $f_\mathrm{car} - f_\mathrm{mod}$, and $f_\mathrm{car} + f_\mathrm{mod}$. Phase locking ensures immunity to phase-dependent artifacts that arise in successive demodulation stages. Moreover, because demodulation occurs in parallel, the method is not limited by the bandwidth of low-pass filters, which typically constrain thermal imaging \cite{zheng_high_2018,alem_how_2021}. As a result, sideband SThM enables faster acquisition of low-frequency thermal signals with enhanced signal-to-noise ratio (SNR). Importantly, this approach is broadly applicable to samples containing active heat sources, whether located at the surface or deeply buried.

Furthermore, the phase of the detected thermal signal reveals the time delay associated with heat propagation from the active device to the SThM probe as illustrated at Fig.~\ref{fig:method}b. Applied to buried structures, this phase-based approach is known as Scanning Thermal Wave Microscopy (STWM) \cite{kwon_scanning_2003}, which enables non-invasive thermal characterization of the environment surrounding self-heating devices embedded deep within a circuit. The measured phase lag can be converted into an apparent heat-wave velocity, determined by the material’s thermal diffusivity $\alpha$, the geometry, and the presence of thermal interfaces along the propagation path, as illustrated at Fig.~\ref{fig:method}b. This provides valuable insight into heat transport mechanisms in complex commercial device architectures. More details about the equipment and experimental parameters can be found in Supplementary Information~1.

\begin{figure}[h!]
    \centering
    \includegraphics[width=0.6\linewidth]{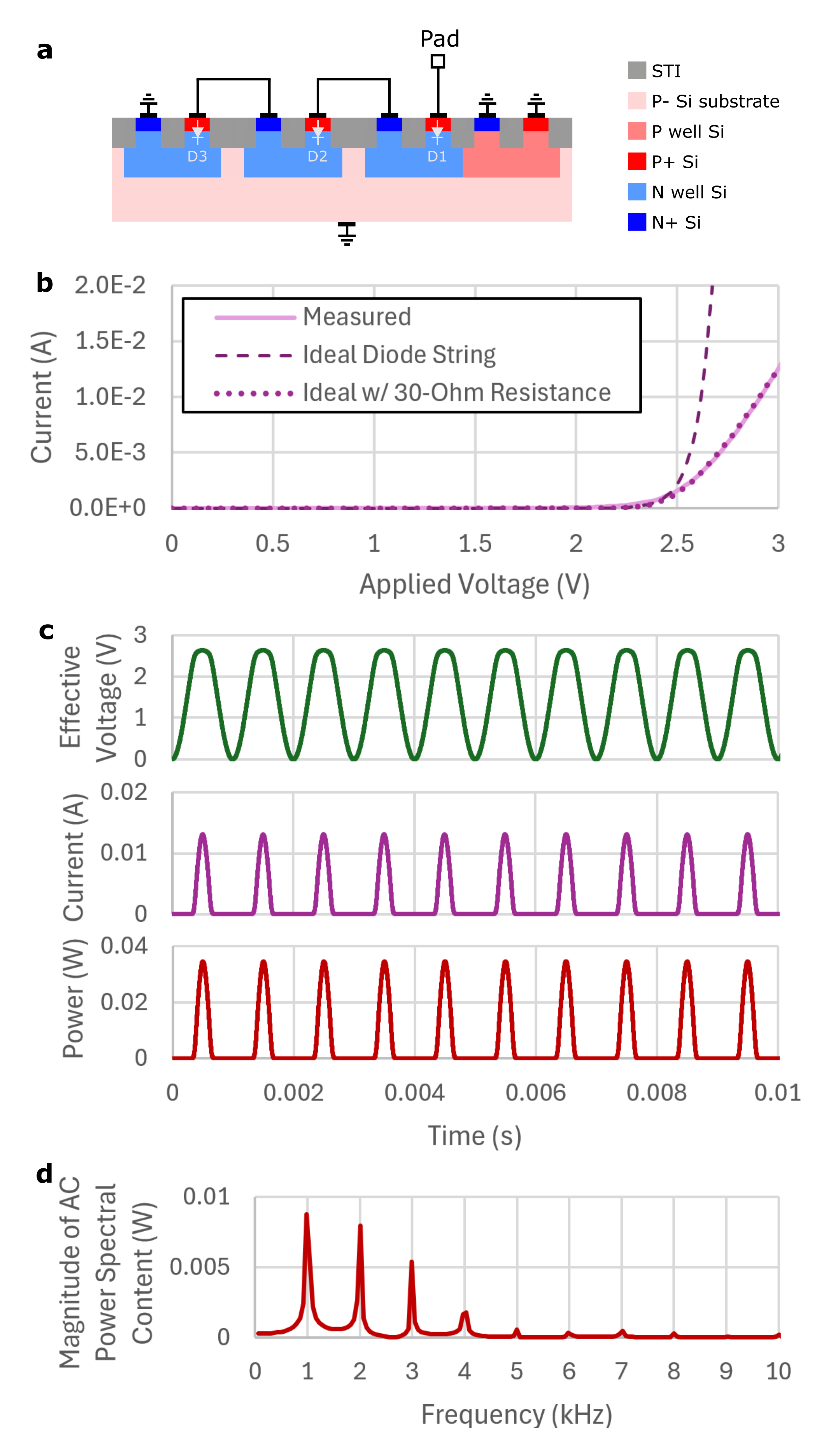}
    \caption{\textbf{a}~Simplified cross-sectional diagram of buried DTSCR heater structure (not to scale) showing diodes D1, D2, and D3. \textbf{b}~Measured DC I-V response of the diode string, which can be modeled as an ideal diode string with 30 $\Omega$ inline series resistance (reducing the effective voltage across the diode string at high bias compared to the sourced voltage). \textbf{c}~Modeled transient waveforms for effective voltage, current, and dissipated power at the diode string. \textbf{d}~Resulting modeled spectral content of power dissipation in the diode string. The fundamental frequency and 2$^{\textrm{nd}}$ harmonic are dominant, though the 3$^{\textrm{rd}}$ harmonic is also significant.}
    \label{fig:dtscr}
\end{figure}

\begin{figure}
    \centering
    \includegraphics[width=0.58\linewidth]{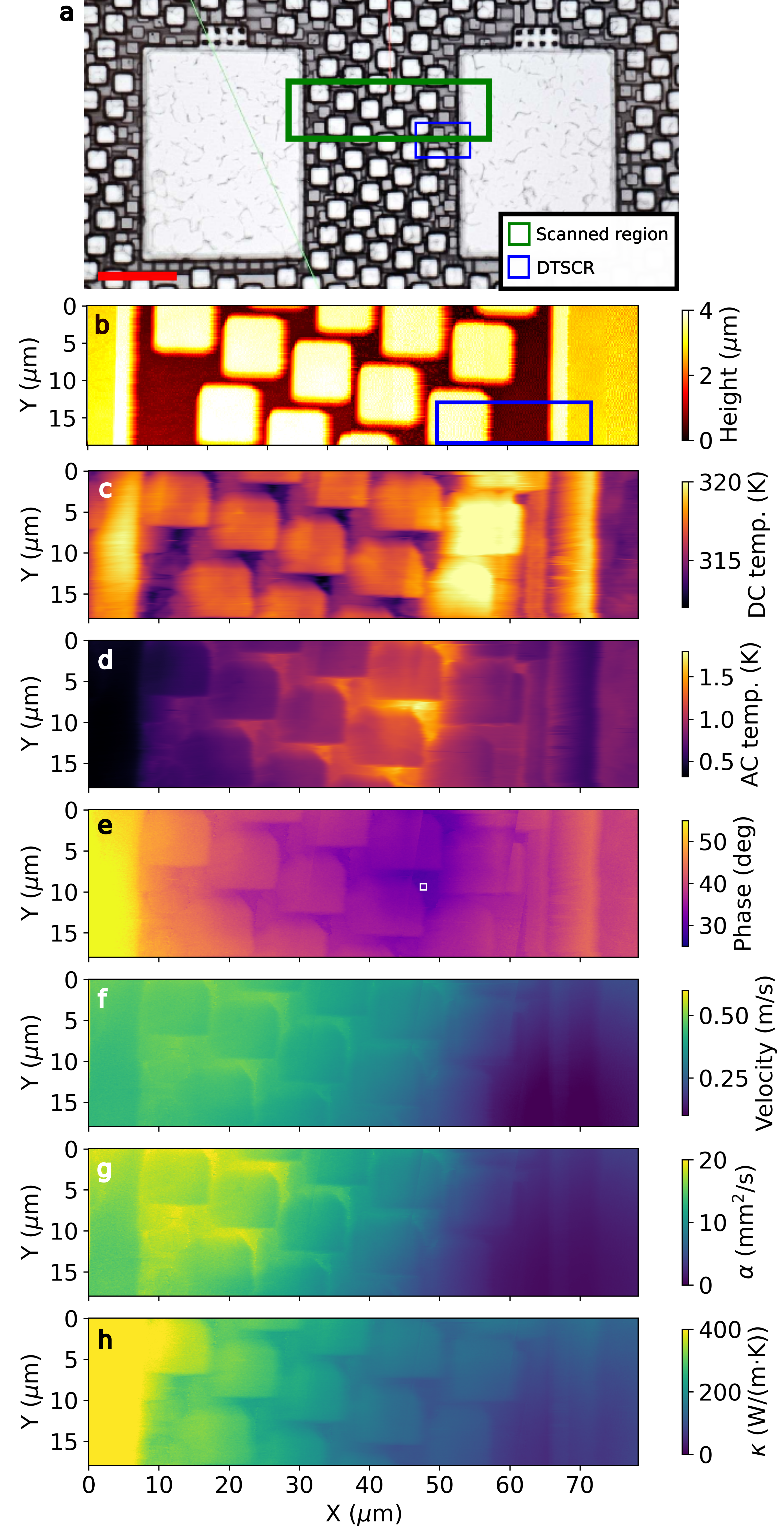}
\caption{\textbf{a}~Grayscale optical micrograph of the region of interest with the scan window indicated by the green rectangle. The buried DTSCR heater is indicated by the blue rectangle. Red scale bar: 30~um. \textbf{b}~Relative topography of the region. \textbf{c}~Apparent DC temperature map. \textbf{d}~Calibrated AC temperature obtained via the sideband detection. \textbf{e}~Relative surface phase lag of the heat wave, obtained through the sideband detection, with the white square indicating the reference region used for the conversion of phase to heat wave velocity. \textbf{f}~Effective heat wave velocity. \textbf{g}~Effective thermal diffusivity. \textbf{h}~Effective thermal conductivity.}
    \label{fig:main_map}
\end{figure}

In this work, the enhanced SNR and increased scanning speed enabled by the sideband SThM technique were leveraged to image surface heating and heat propagation due to a deeply buried heater device in the GlobalFoundries 22-nm FDSOI process. The buried heater device used in this work was a diode-triggered silicon-controlled rectifier (DTSCR) operated in the forward blocking mode such that the effective heating element was a string of 3 vertical silicon diodes connected in series as shown in Fig.~\ref{fig:dtscr}a. The pad noted in Fig.~\ref{fig:dtscr}a is located outside the visible window of Fig.~\ref{fig:main_map}a, and the electrical connection to the DTSCR is made through routing on buried metal layers.

Careful excitation is required to heat the diode string without triggering the SCR or causing avalanche breakdown. For this purpose, a unipolar sinusoidal signal (0–3 V) was applied, with its maximum set between the diode string’s turn-on voltage and the SCR breakover voltage (Fig.~\ref{fig:dtscr}b). This turn-on voltage was confirmed using the diode's self-heating at Supplementary Information~2.

The unipolar sinusoidal excitation of a purely resistive Ohmic element gives rise to a thermal response comprising a DC component, as well as first and second harmonics. Additional harmonics result from the non-Ohmic resistance of the diode string as shown in Fig.~\ref{fig:dtscr}c and d, but these additional higher-order harmonics are non-dominant due to their smaller magnitude and the material attenuation of high-frequency components and are therefore ignored in this analysis. Since the resistance of the SThM probe is monitored at $f_\mathrm{car}$, this thermal modulation results in four major amplitude-modulated sidebands, appearing at frequency offsets of $\pm f_\mathrm{mod}$ and $\pm 2f_\mathrm{mod}$ from the carrier frequency $f_\mathrm{car}$. The demodulated signal at $f_\mathrm{car}$ reflects changes in the DC temperature component of the surface. More details about the model can be found in Supplementary Information~3.

Fig.~\ref{fig:main_map}a presents an optical micrograph of the sample. The surface is predominantly covered by an insulating passivation layer, with the exception of the large Al pads, which remain unpassivated except for a narrow band bordering their perimeters. The measured topography at the boundary between passivated and unpassivated regions of the pads reveals the passivation layer thickness to be apx. $1 \,\unit{\um}$. The region between the Al pads is fully passivated and contains smaller, floating square metallic islands, also formed from the same Al top metal layer (apx. $3 \,\unit{\um}$ thick based on measured surface height differential at pad/island edges). The DTSCR (highlighted by the blue rectangle) is buried at a depth of at least $6.9 \,\unit{\um}$ below the passivation layer where there is no Al top metal \cite{gf2020stackup}. The region between the Al/passivation top layers and the DTSCR consists of several layers of copper routing and insulating dielectric material (see Supporting Information~4 for more information). Owing to the mirror symmetry of the device, only half of the region above the DTSCR was scanned. This approach is advantageous given the limited scanning window of the SThM system, indicated by the green rectangle in Fig.~\ref{fig:main_map}a.

\section{Results and Analysis}
Fig.~\ref{fig:main_map}b presents the relative AFM topography of this region, while Fig.~\ref{fig:main_map}c displays the corresponding DC response of the SThM probe. The strongest DC temperature response is observed in and around the region above the DTSCR. The enhanced signal over the floating metal islands relative to the inter-island areas is likely influenced by a combination of thermal contact variations and topographic artifacts that cannot be fully disentangled from the DC response. In contrast, the reduced response on the unpassivated metal pads can be attributed to their large area and high thermal conductance, which facilitate efficient heat dissipation. 
To obtain the AC temperature map, a calibration procedure is required to extract the tip-sample contact resistance. This is done by leveraging the relation ship between the DC and AC thermal mapping, also known as dual scan SThM. As the diode is intrinsically highly non-linear, we proceed as Harnack \emph{et al.} to obtain the AC temperature map while extracting the contact resistance. 
From the DTSCR's power density measured at Fig.~\ref{fig:dtscr}d  the $\beta_{1\omega}$ factor can be estimated. As in \cite{harnack_scanning_2025}, by assuming that the temperature difference depend linearly on the device self-heating power, the $\beta_{1\omega}$ factor can thus be written as the ratio of the dissipated power in DC and at the first harmonic, obtained from Fig.~\ref{fig:dtscr}. this holds true as long as the thermal system can be assumed to respond similarly in DC and at the first harmonic.

\begin{equation}
    \beta_{1\omega} \ = \ \frac{\Delta T_{\textrm{AC,}1\omega}}{\Delta T_{\textrm{DC}}}  \ \approx\ \frac{P_{\textrm{dev,AC,}1\omega}}{P_{\textrm{dev,DC}}}  \ = \ 0.911
\end{equation}

The AC temperature map can thus be obtained using this reformulation, based on \cite{harnack_scanning_2025}.

\begin{equation}
    \Delta T_{AC,1\omega} \ =\ \Delta T_{\textrm{sens}0} \frac{\beta_{1\omega}\Delta V_{\textrm{AC,}1\omega}}{\Delta V_{\textrm{AC,}1\omega} - \beta_{1\omega}\Delta V_{\textrm{DC}}}
\end{equation}

Using this calibration, the AC temperature map is obtained and amplitude- and offset-dependent studies of the thermal response to DTSCR heating are provided in Figures~S1 and S2. From the data in Fig.~S1, the sensitivity of the method is estimated to be $254~\mu\mathrm{K}/\sqrt{\mathrm{Hz}}$.

\begin{figure}[h!]
    \centering
    \includegraphics[width=0.8\linewidth]{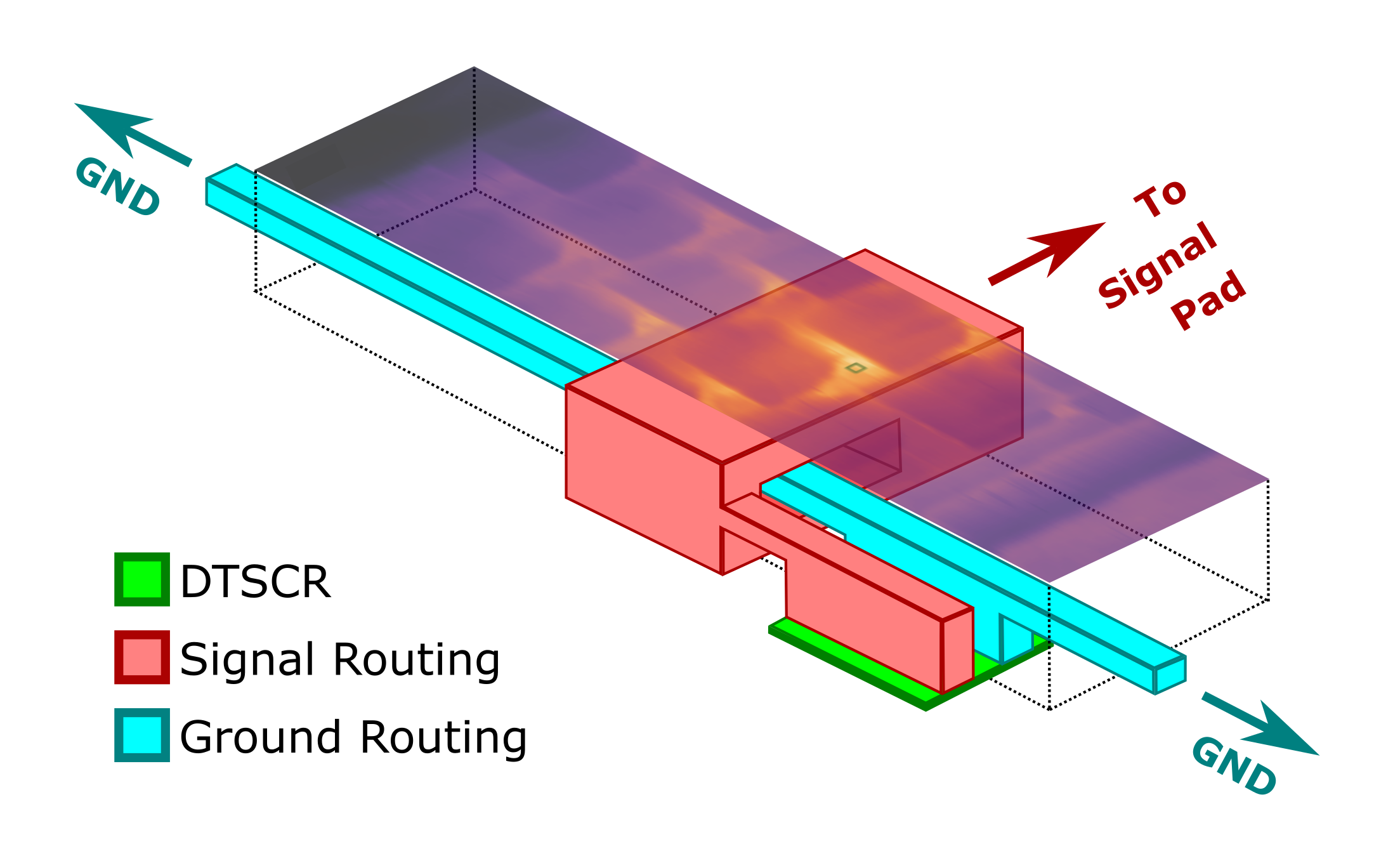}
\caption{Representative 3D routing scheme for DTSCR signal and ground connections (not to scale). The routing regions indicated include both metal routing layers and vias. Metal-island fill and other routing unrelated to the DTSCR are not shown. A partially-transparent copy of Figure \ref{fig:main_map}d is overlaid to demonstrate the maximum AC heating directly above the near-surface signal routing.}
    \label{fig:3d_routing}
\end{figure}

The AC temperature map shows a reduced response above the DTSCR, with the strongest signal near the center of the scan window. This behavior can be attributed to the signal routing: AC heat transfer is most effective in highly conductive, low–specific-heat materials such as metals. As shown in Fig.~\ref{fig:3d_routing}, the signal-routing metal passes close to the chip surface at the scan center, whereas directly above the DTSCR the thick dielectric layer strongly damps the AC component before it reaches the surface. The ground routing above the DTSCR likely contributes further by connecting to a large chip-spanning metal volume with very high thermal capacitance, thereby masking the local AC response. These findings indicate that insulator and/or ground-metal regions can effectively shield temperature-sensitive components from AC fluctuations. Additional measurements on purpose-designed structures will be required to directly quantify this effect for different geometries and frequencies. To avoid artifacts from heat dissipation in the routing itself, the resistance of the routing should be minimized, as implemented here by using large cross-sections (see Fig.~\ref{fig:3d_routing}).

\begin{figure}
    \centering
    \includegraphics[width=0.6\linewidth]{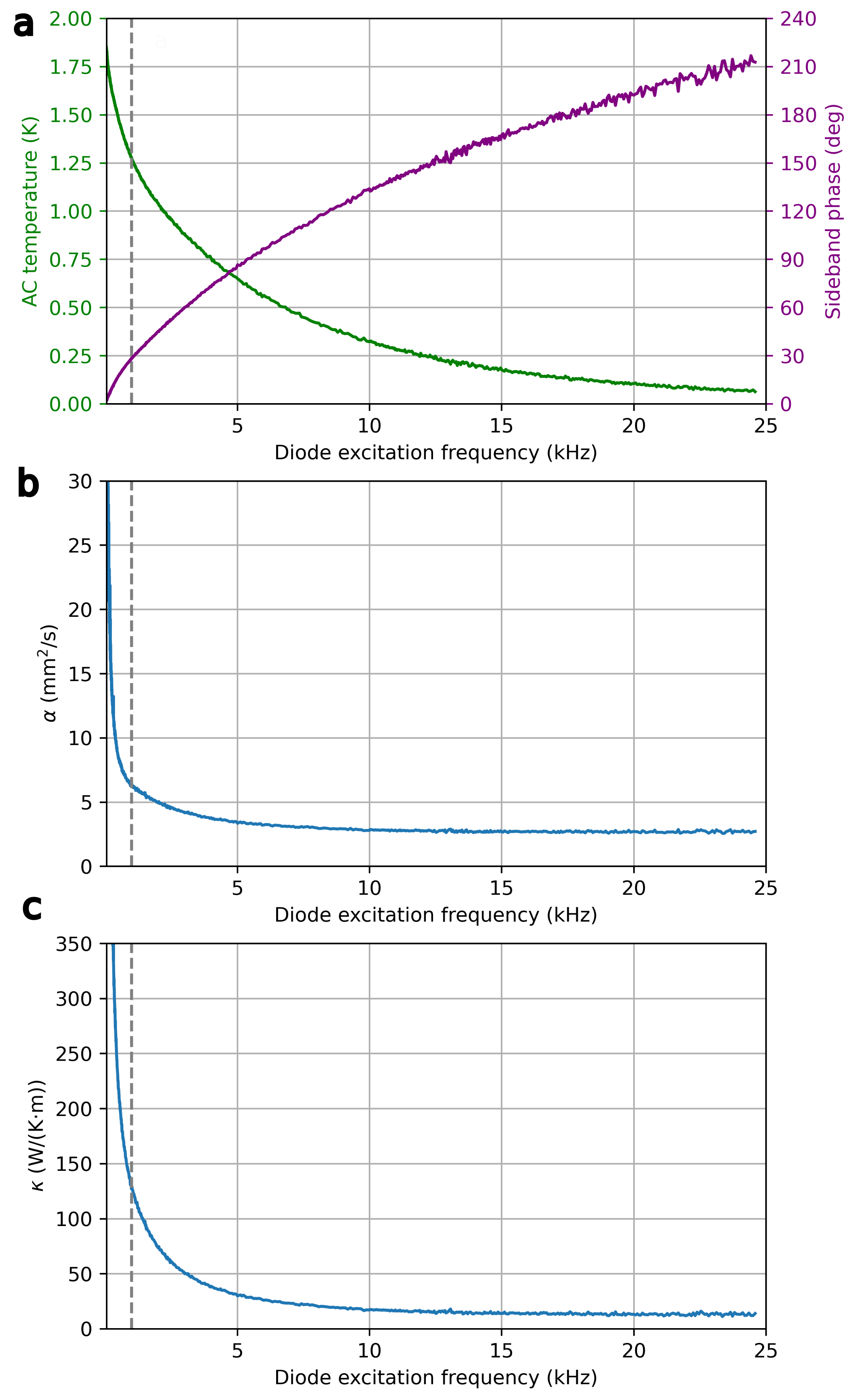}
    \caption{Measured and extracted thermal characteristics at the reference region as a function of the DTSCR's excitation frequency. The excitation is unipolar and its peak-to-peak amplitude is 3~V. The dashed lines highlight the $1$~kHz frequency. \textbf{a} Measured amplitude and phase of the first sideband signal. \textbf{b}~Extracted effective thermal diffusivity. \textbf{c}~Extracted effective thermal conductivity. }
    \label{fig:frequency}
\end{figure}

In the following, we investigate thermal-wave propagation in these devices using our S-STWM approach. Fig.~\ref{fig:main_map}e shows the phase lag between the DTSCR driving signal and the SThM response. The central region of the scan window exhibits both the largest AC heat-wave amplitude and the fastest propagation to the chip surface—exceeding even the region directly above the DTSCR. This behavior arises from the higher heat-wave velocity in the metal routing compared to the surrounding dielectric. The small region marked by a white square in Fig.~\ref{fig:main_map}e was selected for analysis across DTSCR excitation frequencies to enable phase calibration and extraction of effective material properties such as thermal diffusivity $\alpha$ and conductivity $\kappa$. This region's frequency-dependent amplitude and phase responses are shown in Fig.~\ref{fig:frequency}a. The theoretical temperature oscillations $T_{\mathrm{ac}}(x, f)$ at a distance $x$ from a heat source with excitation frequency $f$ in a semi-infinite solid are described by \cite{carslaw_conduction_1959,salazar_energy_2006,kwon_scanning_2003}:
\begin{equation}
|T_{\mathrm{ac}}(x, f)|=\frac{Q_\mathrm{o}}{2 \kappa}\sqrt{\frac{\alpha}{2 \pi f}} \mathrm{e}^{-x/\mu(f)}=\frac{Q_\mathrm{o}}{2 \kappa}\sqrt{\frac{\alpha}{2 \pi f}} \mathrm{e}^{-x\sqrt{\frac{\pi f}{\alpha}}}
\label{eq:wave_ampl}
\end{equation}
\begin{equation}
\phi(x, f)=\frac{2 \pi x}{\lambda(f)}=x\sqrt{\frac{\pi f}{\alpha}}
\label{eq:wave_ph}
\end{equation}
where $Q_0$ is the surface heat flux leaving the heat source; $\kappa$ is the effective thermal conductivity; $\mu(f)=\sqrt{\alpha / \pi f}$ is the penetration depth of the thermal wave; $\phi(x, f)$ is the phase delay in radians of the thermal wave at distance $x$ from the heat source; and $\lambda(f)=2\sqrt{\pi \alpha / f}$ is the effective wavelength of the thermal wave. Analysis of the amplitude data is complicated by its dependency on multiple unknown properties, i.e. $\alpha$ and $\kappa$. Additionally, as $f$ approaches zero, the theoretical amplitude would approach infinity, making Eqn. \ref{eq:wave_ampl} invalid for low frequencies. Phase analysis is more straightforward, relying on fewer assumptions and having only $\alpha$ as an unknown dependency. This allows for direct extraction of effective diffusivity from phase data and the known distance $x$ from the DTSCR. 

Before processing the phase data, an offset calibration is applied to the raw measured relative phase data such that the phase approaches zero as frequency approaches zero to ensure accurate representation of the physical absolute phase delay. At 1 kHz, the corrected phase delay at this reference region is 28.38$^\circ$. The 1 kHz phase map in Fig.~\ref{fig:main_map}e has the same offset applied and is thus an accurate representation of absolute phase delay across the entire region. Since the distance between the contact point and the sensing element is typically below \SI{0.1}{\micro\meter}, any phase delay occurring within the probe can be neglected~\cite{tovee_nanoscale_2012}. To further validate the measured phase, we performed amplitude-dependent SThM traces above the DTSCR heater. The results, presented in Supplementary Information~E, show that the lag of the heat waves, represented by the phase signal of the first and second harmonic sidebands, does not depend on the power dissipated at the DTSCR, as expected. This confirms that the observed phase delays are dominated by the propagation of heatwaves.

The distance $x$ is calculated as the approximate straight-line distance in 3D space from the center of the DTSCR heater. The vertical component of distance is determined by applying an offset to the measured topography in Fig.~\ref{fig:main_map}b such that the vertical distance is \SI{7.9}{\micro\meter} in the region directly above the center of the DTSCR where there is no Al top metal, based on the \SI{6.9}{\micro\meter} copper metal stackup \cite{gf2020stackup} plus the measured \SI{1}{\micro\meter} passivation thickness. For example, the calculated distance at the small reference region is apx. \SI{22}{\micro\meter}. This value likely represents a slight underestimate, as it does not include the thickness of the dielectric layer directly above the top Cu layer (and directly below the Al top metal layer where present). Additional uncertainties arise from the simplified distance calculation due to the distributed and non-uniform nature of the DTSCR heating and the complex geometry of metals and insulators, which may result in non-straight-line dominant conduction paths. Nonetheless, this approach provides a useful approximation of distance for analysis purposes. For example, the effective heat wave velocity can be mapped as shown in Fig.~\ref{fig:main_map}f, where for each point the distance is divided by the propagation time corresponding to the phase delay at that point.
The extracted effective diffusivity is plotted across frequency for the reference region in Fig.~\ref{fig:frequency}b, showing a decrease in $\alpha$ with increasing frequency. Frequency-dependent diffusivity is commonly reported for heterogeneous structures, as parallel thermal pathways are likely to have different thermal wave cut-off frequencies~\cite{koh_freq_2007, vadasz_hetero_2006}. The effective diffusivity $\alpha$ at $f$ = 1 kHz can also be mapped across the scan window as shown in Fig.~\ref{fig:main_map}g. Extracted $\alpha$ values range from $1\times5^{-6} - 2\times10^{-5}~\si{\meter\squared\per\second}$, with greater effective diffusivity observed on the left side of the window where a greater portion of the heat transfer pathway to the surface is formed of the signal routing metal. For comparison, typical values are $\alpha \approx 7\times10^{-7}~\si{\meter\squared\per\second}$ for SiO$_2$ and $\alpha \approx (2\times10^{-5} - 1\times10^{-4})~\si{\meter\squared\per\second}$ for thin-film \ch{Cu} \cite{katsura_thermal_1993,lugo_thermal_2016}. Other on-chip metals and insulators may exhibit significant variation, but extracted $\alpha$ values between those of SiO$_2$ and \ch{Cu} are consistent with thermal paths comprising both materials \cite{yang_thermal_2009,lugo_thermal_2016}. Furthermore, thermal boundary resistance at metal–insulator and insulator–insulator interfaces may also contribute to the effective value of $\alpha$ \cite{he_modelling_2018}. The range of $\alpha$ values extracted demonstrates strong dependence on routing geometry and frequency, with insulator-dominated heat flow in some conditions and metal-dominated heat flow in others. 

With $\alpha$ having been extracted from phase data, the only remaining unknown in Eqn.~\ref{eq:wave_ampl} is $\kappa$, assuming $Q_\mathrm{o}$ can be adequately estimated. The DTSCR power is taken to be 8.75~mW from the model of first-harmonic power in Fig.~\ref{fig:dtscr}d and is assumed constant over frequency. The DTSCR is approximated as a 2D object, and thus its surface area is given by 2 times the area of the region it occupies (see Supplementary Information~4) for a total of $\sim500\,\unit{\um}^2$. This yields $Q_\mathrm{o}\approx1.7\times10^7$~W/m$^2$. Based on this estimate, Eqn.~\ref{eq:wave_ampl} is used to extract $\kappa$ across frequency for the reference region as shown in Fig.~\ref{fig:frequency}c. At the lowest frequencies, the $\kappa$ values extracted appear too high to be physical, exceeding even the conductivity of Cu, which has $\kappa \approx 109-350+$~W/(m$\cdot$K) at room temperature depending on film thickness~\cite{lugo_thermal_2016}. This can likely be attributed to the lack of validity of Eqn.~\ref{eq:wave_ampl} at low frequencies and the added contributions of thermal waves reflected off the chip's bottom surface at low frequencies (when $\mu(f)$ of the substrate material approaches or exceeds two times the thickness of the chip) which violate the assumption of semi-infinite 3D propagation. At the reference region, extracted $\kappa$ at $f$ = 1~kHz is 127~W/(m$\cdot$K), decreasing to as low as 13~W/(m$\cdot$K) with increasing frequency. This reflects well the range of conductivities between those of Cu and SiO$_2$, which has $\kappa \approx 1.1$~W/(m$\cdot$K) at room temperature~\cite{kleiner_sio2_1996}. While some overestimation of $\kappa$ is inherent to the application of this methodology at lower frequencies, decreasing $\kappa$ with increasing frequency is commonly reported for heterogeneous systems and can be attributed to local nonequilibrium effects and reduced penetration depth at higher frequencies leading to the exclusion of contributions from phonons with larger mean free paths~\cite{sobolev_freq_2025, koh_freq_2007}. 

The extracted conductivity $\kappa$ at $f$ = 1~kHz is mapped across the scan window in Fig.~\ref{fig:main_map}h, though extracted values at this frequency may be overestimations. Additionally, Eqn.~\ref{eq:wave_ampl} becomes a worse approximation at greater horizontal distance from the DTSCR, as the solid-air boundary at the chip top surface causes increased deviation in thermal wave propagation compared to the assumed semi-infinite 3D propagation. Higher uncertainty, and therefore error, in extracted $\kappa$ is also expected in regions of low detected amplitude of temperature oscillations, such as the far left pad region. However, two clear patterns emerge: higher effective $\kappa$ in regions above the Al top metal islands compared to their surrounding regions, and higher effective $\kappa$ moving toward the left side of the imaged region. Both of these patterns reflect the higher relative metal content in these regions' paths from the DTSCR, aligning well with expectations.

\begin{figure}[h!]
    \centering
    \includegraphics[width=0.65\linewidth]{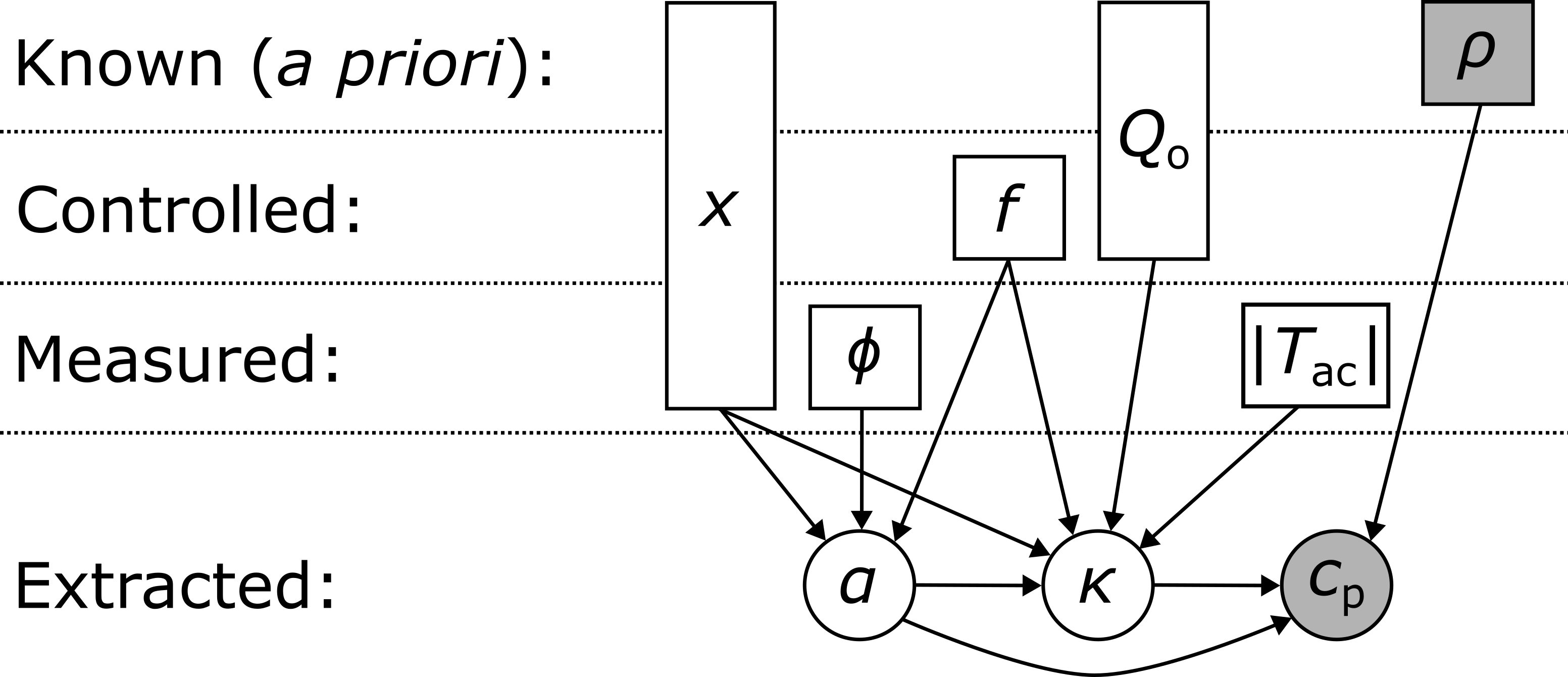}
    \caption{Flowchart of effective thermal material property extraction procedure enabled by S-STWM. Extraction of $c_\mathrm{p}$ is not demonstrated in this work.}
    \label{fig:flow}
\end{figure}

If the effective mass density $\rho$ of the BEOL material could be reasonably estimated, an estimate of the effective specific heat capacity $c_\mathrm{p}=\kappa / \alpha\rho $ could be derived as well. A flowchart of the full procedure of thermal material property extraction is shown in Fig.~\ref{fig:flow}. This highlights that as the extraction moves from $\alpha$ to $\kappa$ to $c_\mathrm{p}$, additional error sources are added at each step, whether from measurement uncertainty, estimations of \textit{a priori} information, or invalid model assumptions. Extraction of $c_\mathrm{p}$ is not carried out in this work since the complexity of the structure does not allow for a good estimation of effective $\rho$, but future studies of purpose-designed regions of simpler BEOL composition above a buried heater element may allow for good $\rho$ estimation and useful extraction of effective $c_\mathrm{p}$. Measuring several variants of such structures, especially across a range of cryogenic ambient temperatures, could then allow for extraction of individual contributions of different BEOL layers/boundaries and estimates of their independent effective material properties. This would then enable predictive 3D finite element analysis (FEA) simulation capabilities, including transient effects up to the characterized frequency range.

\section{Conclusion}
Realizing the full potential of quantum computing requires seamless integration of qubit architectures with cryogenic electronics, which in turn demands a detailed understanding of heat propagation within complex 3D circuit structures. Here, we present a novel implementation of AC-Scanning Thermal Microscopy combined with thermal wave analysis and sideband detection, S-STWM, to non-invasively characterize the thermal properties of materials and pathways in the BEOL of a CMOS integrated circuit, spanning from the chip surface to active devices buried within multilayer structures. Our approach provides valuable insight into heat transport in structures where interfacial effects dominate, revealing details that are otherwise inaccessible with conventional methods. 

While the measurements reported here were performed at room temperature, they establish a foundation for equivalent studies at cryogenic conditions~\cite{Fonck_2025}, directly relevant to the operating environment of cryo-CMOS devices. Beyond cryo-electronics applications, the rapid development and adoption of ultra-high-power-density CMOS CPUs and GPUs for high-performance compute (HPC) applications presents an additional possible use case for this methodology \cite{abo-zahhad_thermal_2022,yu_foundry_2021} to augment thermal characterization capabilities of such devices and enable improved thermally-aware design and packaging.\\

\section*{Acknowledgment}

The authors thank Alberto Gomez-Saiz of Quantum Motion for IC design and layout of the DTSCR structure, and Martin Vanbrabant and Jean-Pierre Raskin of UCLouvain for acquiring and providing the micrograph in Figure \ref{fig:main_map}a. 

The authors acknowledge financial support from the F.R.S.-FNRS of Belgium (FNRS-CQ-1.C044.21-SMARD, FNRS-MIS-F.4523.22-TopoBrain, FNRS-PDR-T.0128.24-ART-MULTI, FNRS-CR-1.B.463.22-MouleFrits, FNRS-FRIA-1.E092.23-TOTEM), from the EU (ERC-StG-10104144-MOUNTAIN), from the Federation Wallonie-Bruxelles through the ARC Grant No. 21/26-116, from the FWO and FRS-FNRS under the Excellence of Science (EOS) programme (40007563-CONNECT), and from the UK Industrial Strategy Challenge Fund (ISCF)
Project Altnaharra (Innovate U.K. Grant 10006186).

\section*{Author Contributions}

V. Fonck and M. Razeghi acquired the data. V. Fonck, M. Razeghi, G. Ridgard, G. M. Noah, and P. Gehring analyzed the data. G. M. Noah and P. Gehring packaged the chip. V. Fonck, G. Ridgard, G. M. Noah, and P. Gehring conceived the experiments. V. Fonck and G. M. Noah conceived the effective thermal material property extraction procedure. P. Dobson and J. Weaver fabricated the SThM probes. G. M. Noah and P. Gehring supervised the work. All authors contributed to the writing of this article.

\newpage
\bibliographystyle{IEEEtran}

\bibliography{mybib}

\begin{thebibliography}{10}
\providecommand{\url}[1]{#1}
\csname url@samestyle\endcsname
\providecommand{\newblock}{\relax}
\providecommand{\bibinfo}[2]{#2}
\providecommand{\BIBentrySTDinterwordspacing}{\spaceskip=0pt\relax}
\providecommand{\BIBentryALTinterwordstretchfactor}{4}
\providecommand{\BIBentryALTinterwordspacing}{\spaceskip=\fontdimen2\font plus
\BIBentryALTinterwordstretchfactor\fontdimen3\font minus \fontdimen4\font\relax}
\providecommand{\BIBforeignlanguage}[2]{{%
\expandafter\ifx\csname l@#1\endcsname\relax
\typeout{** WARNING: IEEEtran.bst: No hyphenation pattern has been}%
\typeout{** loaded for the language `#1'. Using the pattern for}%
\typeout{** the default language instead.}%
\else
\language=\csname l@#1\endcsname
\fi
#2}}
\providecommand{\BIBdecl}{\relax}
\BIBdecl

\bibitem{abughanem_ibm_2025}
M.~AbuGhanem, ``{IBM} quantum computers: evolution, performance, and future directions,'' \emph{The Journal of Supercomputing}, vol.~81, Apr. 2025.

\bibitem{acharya2022overcoming}
R.~Acharya, S.~Brebels, A.~Grill, J.~Verjauw, T.~Ivanov, D.~P. Lozano, D.~Wan, J.~Van~Damme, A.~Vadiraj, M.~Mongillo \emph{et~al.}, ``Overcoming i/o bottleneck in superconducting quantum computing: multiplexed qubit control with ultra-low-power, base-temperature cryo-cmos multiplexer,'' \emph{arXiv preprint arXiv:2209.13060}, 2022.

\bibitem{JvS2025snh}
J.~v. Staveren, L.~Enthoven, P.~L. Bavdaz, M.~Meyer, C.~Deprez, V.~Nuutinen, R.~Lake, D.~D. Esposti, C.~Carlsson, A.~Tosato, J.~Gong, B.~Prabowo, M.~Babaie, C.~G. Almudever, M.~Veldhorst, G.~Scappucci, and F.~Sebastiano, ``{ Cryo-CMOS Bias-Voltage Generation and Demultiplexing at mK Temperatures for Large-Scale Arrays of Quantum Devices },'' \emph{IEEE Transactions on Quantum Engineering}, no.~01, pp. 1--19, Jun. 2025.

\bibitem{pauka2021cryocmos}
S.~J. Pauka, K.~Das, R.~Kalra, A.~Moini, Y.~Yang, M.~Trainer, A.~Bousquet, C.~Cantaloube, N.~Dick, G.~C. Gardner, M.~J. Manfra, and D.~J. Reilly, ``A cryogenic cmos chip for generating control signals for multiple qubits.'' \emph{Nature Electronics}, vol.~4, pp. 64--70, 2021.

\bibitem{intel2024pandotree}
S.~Subramanian, T.~M. Mladenov, S.~Schaal, B.~Patra, L.~Lampert, N.~K. Robinson, J.~Roberts, and S.~Pellerano, ``A scalable mk cryo-cmos demultiplexer chip for voltage biasing and high-speed control of silicon qubit gates,'' in \emph{2024 IEEE Symposium on VLSI Technology and Circuits (VLSI Technology and Circuits)}, 2024, pp. 1--2.

\bibitem{de_kruijf_measurement_2024}
M.~de~Kruijf, G.~M. Noah, A.~Gomez-Saiz, J.~J.~L. Morton, and M.~F. Gonzalez-Zalba, ``Measurement of cryoelectronics heating using a local quantum dot thermometer in silicon,'' \emph{Chip}, vol.~3, no.~3, p. 100097, Sep. 2024.

\bibitem{noah_cmos_2024}
G.~M. Noah, T.~H. Swift, M.~de~Kruijf, A.~Gomez-Saiz, J.~J.~L. Morton, and M.~F. Gonzalez-Zalba, ``{CMOS} on-chip thermometry at deep cryogenic temperatures,'' \emph{Applied Physics Reviews}, vol.~11, no.~2, p. 021414, Apr. 2024.

\bibitem{t_hart_characterization_2021}
P.~A. T~Hart, M.~Babaie, A.~Vladimirescu, and F.~Sebastiano, ``Characterization and {Modeling} of {Self}-{Heating} in {Nanometer} {Bulk}-{CMOS} at {Cryogenic} {Temperatures},'' \emph{IEEE Journal of the Electron Devices Society}, vol.~9, pp. 891--901, 2021.

\bibitem{huang_qubit_2024}
\BIBentryALTinterwordspacing
J.~Y. Huang, R.~Y. Su, W.~H. Lim, M.~Feng, B.~van Straaten, B.~Severin, W.~Gilbert, N.~D. Stuyck, T.~Tanttu, S.~Serrano \emph{et~al.}, ``High-fidelity spin qubit operation and algorithmic initialization above 1 k,'' \emph{Nature}, vol. 623, pp. 772--777, 2024. [Online]. Available: \url{https://doi.org/10.1038/s41586-024-07160-2}
\BIBentrySTDinterwordspacing

\bibitem{Fonck_2025}
\BIBentryALTinterwordspacing
V.~Fonck, J.~Spiece, and P.~Gehring, ``Quantum heat under the microscope: a perspective on cryogenic scanning thermal microscopy,'' \emph{Nano Futures}, vol.~9, no.~3, p. 032502, jul 2025. [Online]. Available: \url{https://dx.doi.org/10.1088/2399-1984/adef25}
\BIBentrySTDinterwordspacing

\bibitem{woon_thermal_2025}
W.-Y. Woon, A.~Kasperovich, J.-R. Wen, K.~K. Hu, M.~Malakoutian, J.-H. Jhang, S.~Vaziri, I.~Datye, C.~C. Shih, J.~F. Hsu, X.~Y. Bao, Y.~Wu, M.~Nomura, S.~Chowdhury, and S.~S. Liao, ``\BIBforeignlanguage{en}{Thermal management materials for {3D}-stacked integrated circuits},'' \emph{\BIBforeignlanguage{en}{Nature Reviews Electrical Engineering}}, pp. 1--16, Aug. 2025.

\bibitem{thompson1961thermal}
J.~Thompson and B.~Younglove, ``Thermal conductivity of silicon at low temperatures,'' \emph{Journal of Physics and Chemistry of Solids}, vol.~20, no. 1-2, pp. 146--149, 1961.

\bibitem{pohl2002low}
R.~O. Pohl, X.~Liu, and E.~Thompson, ``Low-temperature thermal conductivity and acoustic attenuation in amorphous solids,'' \emph{Reviews of Modern Physics}, vol.~74, no.~4, p. 991, 2002.

\bibitem{swartz1989thermal}
E.~T. Swartz and R.~O. Pohl, ``Thermal boundary resistance,'' \emph{Reviews of modern physics}, vol.~61, no.~3, p. 605, 1989.

\bibitem{ekin_experimental_2007}
J.~Ekin and G.~Zimmerman, \emph{Experimental {Techniques} for {Low}-{Temperature} {Measurements}}.\hskip 1em plus 0.5em minus 0.4em\relax Oxford University Press, May 2007.

\bibitem{pop2010energy}
E.~Pop, ``Energy dissipation and transport in nanoscale devices,'' \emph{Nano Research}, vol.~3, pp. 147--169, 2010.

\bibitem{artanov2022self}
A.~A. Artanov, E.~A. Guti{\'e}rrez-D, A.~R. Cabrera-Galicia, A.~Kruth, C.~Degenhardt, D.~Durini, J.~M{\'e}ndez-V, and S.~Van~Waasen, ``Self-heating effect in a 65 nm mosfet at cryogenic temperatures,'' \emph{IEEE transactions on electron devices}, vol.~69, no.~3, pp. 900--904, 2022.

\bibitem{pop_heat_2006}
\BIBentryALTinterwordspacing
E.~Pop, S.~Sinha, and K.~Goodson, ``Heat {Generation} and {Transport} in {Nanometer}-{Scale} {Transistors},'' \emph{Proceedings of the IEEE}, vol.~94, no.~8, pp. 1587--1601, Aug. 2006. [Online]. Available: \url{https://ieeexplore.ieee.org/abstract/document/1705144}
\BIBentrySTDinterwordspacing

\bibitem{shiomi2024scientific}
J.~Shiomi and K.~Uchida, ``Scientific challenges of cryo-electronics thermal management,'' \emph{Nature Reviews Electrical Engineering}, vol.~1, no.~12, pp. 762--763, 2024.

\bibitem{Feng_TSV}
\BIBentryALTinterwordspacing
W.~Feng, K.~Kikuchi, M.~Hidaka, H.~Yamamori, Y.~Araga, K.~Makise, and S.~Kawabata, ``Thermal management of a 3d packaging structure for superconducting quantum annealing machines,'' \emph{Applied Physics Letters}, vol. 118, no.~17, p. 174004, 04 2021. [Online]. Available: \url{https://doi.org/10.1063/5.0039822}
\BIBentrySTDinterwordspacing

\bibitem{halder2023beolheatsink}
\BIBentryALTinterwordspacing
A.~Halder, L.~Nyssens, D.~Lederer, V.~Kilchytska, and J.-P. Raskin, ``Comparison of heat sinks in back-end of line to reduce self-heating in 22fdx® mosfets,'' \emph{Solid-State Electronics}, vol. 207, p. 108706, 2023. [Online]. Available: \url{https://www.sciencedirect.com/science/article/pii/S0038110123001193}
\BIBentrySTDinterwordspacing

\bibitem{thomas2022interposer}
\BIBentryALTinterwordspacing
C.~Thomas, J.-P. Michel, E.~Deschaseaux, J.~Charbonnier, R.~Souil, E.~Vermande, A.~Campo, T.~Farjot, G.~Rodriguez, G.~Romano, F.~Gustavo, B.~Jadot, V.~Thiney, Y.~Thonnart, G.~Billiot, T.~Meunier, and M.~Vinet, ``Superconducting routing platform for large-scale integration of quantum technologies,'' \emph{Materials for Quantum Technology}, vol.~2, no.~3, p. 035001, aug 2022. [Online]. Available: \url{https://dx.doi.org/10.1088/2633-4356/ac88ae}
\BIBentrySTDinterwordspacing

\bibitem{swami_experimental_2024}
\BIBentryALTinterwordspacing
R.~Swami, G.~Julié, S.~Le-Denmat, G.~Pernot, D.~Singhal, J.~Paterson, J.~Maire, J.~F. Motte, N.~Paillet, H.~Guillou, S.~Gomès, and O.~Bourgeois, ``Experimental setup for thermal measurements at the nanoscale using a {SThM} probe with niobium nitride thermometer,'' \emph{Review of Scientific Instruments}, vol.~95, no.~5, p. 054904, May 2024. [Online]. Available: \url{https://doi.org/10.1063/5.0203890}
\BIBentrySTDinterwordspacing

\bibitem{bourgeois_liquid_2006}
\BIBentryALTinterwordspacing
O.~Bourgeois, E.~André, C.~Macovei, and J.~Chaussy, ``Liquid nitrogen to room-temperature thermometry using niobium nitride thin films,'' \emph{Review of Scientific Instruments}, vol.~77, no.~12, p. 126108, Dec. 2006. [Online]. Available: \url{https://doi.org/10.1063/1.2403934}
\BIBentrySTDinterwordspacing

\bibitem{menges_quantitative_2012}
F.~Menges, H.~Riel, A.~Stemmer, and B.~Gotsmann, ``Quantitative {Thermometry} of {Nanoscale} {Hot} {Spots},'' \emph{Nano Letters}, vol.~12, no.~2, pp. 596--601, Feb. 2012.

\bibitem{kim_quantitative_2011}
K.~Kim, J.~Chung, G.~Hwang, O.~Kwon, and J.~S. Lee, ``Quantitative {Measurement} with {Scanning} {Thermal} {Microscope} by {Preventing} the {Distortion} {Due} to the {Heat} {Transfer} through the {Air},'' \emph{ACS Nano}, vol.~5, no.~11, pp. 8700--8709, Nov. 2011.

\bibitem{kim_ultra-high_2012}
K.~Kim, W.~Jeong, W.~Lee, and P.~Reddy, ``Ultra-{High} {Vacuum} {Scanning} {Thermal} {Microscopy} for {Nanometer} {Resolution} {Quantitative} {Thermometry},'' \emph{ACS Nano}, vol.~6, no.~5, pp. 4248--4257, May 2012.

\bibitem{saidi_scanning_2009}
E.~Saïdi, B.~Samson, L.~Aigouy, S.~Volz, P.~Löw, C.~Bergaud, and M.~Mortier, ``\BIBforeignlanguage{en}{Scanning thermal imaging by near-field fluorescence spectroscopy},'' \emph{\BIBforeignlanguage{en}{Nanotechnology}}, vol.~20, no.~11, p. 115703, Feb. 2009.

\bibitem{saidi_tuning_2011}
E.~Saïdi, N.~Babinet, L.~Lalouat, J.~Lesueur, L.~Aigouy, S.~Volz, J.~Labéguerie-Egéa, and M.~Mortier, ``\BIBforeignlanguage{en}{Tuning {Temperature} and {Size} of {Hot} {Spots} and {Hot}-{Spot} {Arrays}},'' \emph{\BIBforeignlanguage{en}{Small}}, vol.~7, no.~2, pp. 259--264, 2011.

\bibitem{shi_thermal_2001}
L.~Shi and A.~Majumdar, ``Thermal {Transport} {Mechanisms} at {Nanoscale} {Point} {Contacts},'' \emph{Journal of Heat Transfer}, vol. 124, no.~2, pp. 329--337, Jul. 2001.

\bibitem{fiege_failure_1998}
G.~B.~M. Fiege, V.~Feige, J.~C.~H. Phang, M.~Maywald, S.~Gorlich, and L.~J. Balk, ``Failure analysis of integrated devices by scanning thermal microscopy ({SThM}),'' \emph{Microelectronics Reliability}, vol.~38, no.~6, pp. 957--961, Jun. 1998.

\bibitem{fiege_thermal_1999}
G.~B.~M. Fiege, F.~J. Niedernostheide, H.~J. Schulze, R.~Barthelmeß, and L.~J. Balk, ``Thermal characterization of power devices by scanning thermal microscopy techniques,'' \emph{Microelectronics Reliability}, vol.~39, no.~6, pp. 1149--1152, Jun. 1999.

\bibitem{gomes_temperature_2007}
\BIBentryALTinterwordspacing
S.~Gomes, P.-O. Chapuis, F.~Nepveu, N.~Trannoy, S.~Volz, B.~Charlot, G.~Tessier, S.~Dilhaire, B.~Cretin, and P.~Vairac, ``Temperature {Study} of {Sub}-{Micrometric} {ICs} by {Scanning} {Thermal} {Microscopy},'' \emph{IEEE Transactions on Components and Packaging Technologies}, vol.~30, no.~3, pp. 424--431, Sep. 2007. [Online]. Available: \url{https://ieeexplore.ieee.org/abstract/document/4295156}
\BIBentrySTDinterwordspacing

\bibitem{Menges2016}
\BIBentryALTinterwordspacing
F.~Menges, P.~Mensch, H.~Schmid, H.~Riel, A.~Stemmer, and B.~Gotsmann, ``Temperature mapping of operating nanoscale devices by scanning probe thermometry,'' \emph{Nature Communications}, vol.~7, no.~1, p. 10874, Mar 2016. [Online]. Available: \url{https://doi.org/10.1038/ncomms10874}
\BIBentrySTDinterwordspacing

\bibitem{Harzheim2018}
A.~Harzheim, J.~Spiece, C.~Evangeli, E.~McCann, V.~Falko, Y.~Sheng, J.~H. Warner, G.~A.~D. Briggs, J.~A. Mol, P.~Gehring, and O.~V. Kolosov, ``Geometrically enhanced thermoelectric effects in graphene nanoconstrictions,'' \emph{Nano Letters}, vol.~18, no.~12, pp. 7719--7725, 2018.

\bibitem{zheng_high_2018}
Z.~Zheng, R.~Xu, S.~Ye, S.~Hussain, W.~Ji, P.~Cheng, Y.~Li, Y.~Sugawara, and Z.~Cheng, ``\BIBforeignlanguage{en}{High harmonic exploring on different materials in dynamic atomic force microscopy},'' \emph{\BIBforeignlanguage{en}{Science China Technological Sciences}}, vol.~61, no.~3, pp. 446--452, Mar. 2018.

\bibitem{alem_how_2021}
M.~Alem, ``How to {Demodulate} {Multi}-frequency {Signals} such as {AM}, {FM} and {PM} {\textbar} {Zurich} {Instruments},'' Jan. 2021.

\bibitem{kwon_scanning_2003}
O.~Kwon, L.~Shi, and A.~Majumdar, ``Scanning {Thermal} {Wave} {Microscopy} ({STWM}),'' \emph{Journal of Heat Transfer}, vol. 125, no.~1, pp. 156--163, Jan. 2003.

\bibitem{gf2020stackup}
N.~Wu, ``22fdsoi technology for fully-integrated mmw and radar applications,'' in \emph{2020 23rd International Microwave and Radar Conference (MIKON)}, 2020.

\bibitem{harnack_scanning_2025}
N.~Harnack, S.~Rodehutskors, and B.~Gotsmann, ``Scanning {Thermal} {Microscopy} {Method} for {Self}-{Heating} in {Nonlinear} {Devices} and {Application} to {Filamentary} {Resistive} {Random}-{Access} {Memory},'' \emph{ACS Nano}, vol.~19, no.~5, pp. 5342--5352, Feb. 2025.

\bibitem{carslaw_conduction_1959}
H.~S. Carslaw and J.~C. Jaeger, \emph{Conduction of Heat in Solids}.\hskip 1em plus 0.5em minus 0.4em\relax Oxford: Oxford University Press, 1959.

\bibitem{salazar_energy_2006}
\BIBentryALTinterwordspacing
A.~Salazar, ``\BIBforeignlanguage{en}{Energy propagation of thermal waves},'' \emph{\BIBforeignlanguage{en}{European Journal of Physics}}, vol.~27, no.~6, p. 1349, Sep. 2006. [Online]. Available: \url{https://dx.doi.org/10.1088/0143-0807/27/6/009}
\BIBentrySTDinterwordspacing

\bibitem{tovee_nanoscale_2012}
\BIBentryALTinterwordspacing
P.~Tovee, M.~Pumarol, D.~Zeze, K.~Kjoller, and O.~Kolosov, ``Nanoscale spatial resolution probes for scanning thermal microscopy of solid state materials,'' \emph{Journal of Applied Physics}, vol. 112, no.~11, p. 114317, Dec. 2012. [Online]. Available: \url{https://doi.org/10.1063/1.4767923}
\BIBentrySTDinterwordspacing

\bibitem{koh_freq_2007}
Y.~K. Koh and D.~G. Cahill, ``Frequency dependence of the thermal conductivity of semiconductor alloys,'' \emph{Physical Review B}, vol.~76, p. 075207, Aug. 2007.

\bibitem{vadasz_hetero_2006}
P.~Vadasz, ``Exclusion of oscillations in heterogeneous and bi-composite media thermal conduction,'' \emph{International Journal of Heat and Mass Transfer}, vol.~49, no.~25, pp. 4886--4892, 2006.

\bibitem{katsura_thermal_1993}
\BIBentryALTinterwordspacing
T.~Katsura, ``\BIBforeignlanguage{en}{Thermal diffusivity of silica glass at pressures up to 9 {GPa}},'' \emph{\BIBforeignlanguage{en}{Physics and Chemistry of Minerals}}, vol.~20, no.~3, pp. 201--208, Aug. 1993. [Online]. Available: \url{https://doi.org/10.1007/BF00200122}
\BIBentrySTDinterwordspacing

\bibitem{lugo_thermal_2016}
\BIBentryALTinterwordspacing
J.~M. Lugo and A.~I. Oliva, ``\BIBforeignlanguage{en}{Thermal {Diffusivity} and {Thermal} {Conductivity} of {Copper} {Thin} {Films} at {Ambient} {Conditions}},'' \emph{\BIBforeignlanguage{en}{Journal of Thermophysics and Heat Transfer}}, Jun. 2016. [Online]. Available: \url{https://arc.aiaa.org/doi/10.2514/1.T4727}
\BIBentrySTDinterwordspacing

\bibitem{yang_thermal_2009}
\BIBentryALTinterwordspacing
S.~T. Yang, M.~J. Matthews, S.~Elhadj, V.~G. Draggoo, and S.~E. Bisson, ``Thermal transport in {CO2} laser irradiated fused silica: {In} situ measurements and analysis,'' \emph{Journal of Applied Physics}, vol. 106, no.~10, p. 103106, Nov. 2009. [Online]. Available: \url{https://doi.org/10.1063/1.3259419}
\BIBentrySTDinterwordspacing

\bibitem{he_modelling_2018}
\BIBentryALTinterwordspacing
J.~He, Q.~Liu, Z.~Wu, and X.~Xu, ``Modelling transient heat conduction of granular materials by numerical manifold method,'' \emph{Engineering Analysis with Boundary Elements}, vol.~86, pp. 45--55, Jan. 2018. [Online]. Available: \url{https://doi.org/10.1016/j.enganabound.2017.10.011}
\BIBentrySTDinterwordspacing

\bibitem{kleiner_sio2_1996}
M.~Kleiner, S.~Kuhn, and W.~Weber, ``Thermal conductivity measurements of thin silicon dioxide films in integrated circuits,'' \emph{IEEE Transactions on Electron Devices}, vol.~43, no.~9, pp. 1602--1609, 1996.

\bibitem{sobolev_freq_2025}
\BIBentryALTinterwordspacing
S.~Sobolev and I.~Kudinov, ``Dispersion relation and frequency-dependent thermal conductivity of the two-temperature systems,'' \emph{International Journal of Thermal Sciences}, vol. 214, p. 109937, Aug. 2025. [Online]. Available: \url{https://doi.org/10.1016/j.ijthermalsci.2025.109937}
\BIBentrySTDinterwordspacing

\bibitem{abo-zahhad_thermal_2022}
E.~M. Abo-Zahhad, A.~Amine~Hachicha, Z.~Said, C.~Ghenai, and S.~Ookawara, ``Thermal management system for high, dense, and compact power electronics,'' \emph{Energy Conversion and Management}, vol. 268, p. 115975, Sep. 2022.

\bibitem{yu_foundry_2021}
D.~C.~H. Yu, C.-T. Wang, and H.~Hsia, ``Foundry {Perspectives} on 2.{5D}/{3D} {Integration} and {Roadmap},'' in \emph{2021 {IEEE} {International} {Electron} {Devices} {Meeting} ({IEDM})}, Dec. 2021, pp. 3.7.1--3.7.4.

\end{thebibliography}


\begin{thebibliography}{1}
\providecommand{\url}[1]{#1}
\csname url@samestyle\endcsname
\providecommand{\newblock}{\relax}
\providecommand{\bibinfo}[2]{#2}
\providecommand{\BIBentrySTDinterwordspacing}{\spaceskip=0pt\relax}
\providecommand{\BIBentryALTinterwordstretchfactor}{4}
\providecommand{\BIBentryALTinterwordspacing}{\spaceskip=\fontdimen2\font plus
\BIBentryALTinterwordstretchfactor\fontdimen3\font minus \fontdimen4\font\relax}
\providecommand{\BIBforeignlanguage}[2]{{%
\expandafter\ifx\csname l@#1\endcsname\relax
\typeout{** WARNING: IEEEtran.bst: No hyphenation pattern has been}%
\typeout{** loaded for the language `#1'. Using the pattern for}%
\typeout{** the default language instead.}%
\else
\language=\csname l@#1\endcsname
\fi
#2}}
\providecommand{\BIBdecl}{\relax}
\BIBdecl

\bibitem{dobson_new_2007}
\BIBentryALTinterwordspacing
P.~S. Dobson, J.~M.~R. Weaver, and G.~Mills, ``New {Methods} for {Calibrated} {Scanning} {Thermal} {Microscopy} ({SThM}),'' in \emph{2007 {IEEE} {SENSORS}}, Oct. 2007, pp. 708--711. [Online]. Available: \url{https://ieeexplore.ieee.org/abstract/document/4388498}
\BIBentrySTDinterwordspacing

\bibitem{klewer2019qual}
C.~Klewer, F.~Kuechenmeister, J.~Paul, D.~Breuer, B.~Boehme, J.~K. Cho, S.~Capecchi, and M.~Thiele, ``Package {Qualification} {Envelope} for 22fdx\textregistered{} {Technology},'' \emph{IMAPSource Proceedings}, vol. 2019, no.~1, pp. 169--175, oct 1 2019.

\bibitem{bulla1998teos}
\BIBentryALTinterwordspacing
D.~Bulla and N.~Morimoto, ``Deposition of thick teos pecvd silicon oxide layers for integrated optical waveguide applications,'' \emph{Thin Solid Films}, vol. 334, no.~1, pp. 60--64, 1998. [Online]. Available: \url{https://www.sciencedirect.com/science/article/pii/S0040609098011171}
\BIBentrySTDinterwordspacing

\bibitem{gf2020stackup}
N.~Wu, ``22fdsoi technology for fully-integrated mmw and radar applications,'' in \emph{2020 23rd International Microwave and Radar Conference (MIKON)}, 2020.

\end{thebibliography}

\end{document}


\section{Supplementary information for "Characterization of heat transfer in 3D CMOS structures using Sideband Scanning Thermal Wave Microscopy"}
\section*{Supporting Information}

\subsection{Experimental details}
\label{sup:exp_details}
This work was realised on a modified Thorlabs Atomic Force Microscope (Thorlabs Inc). The probe used consists of a triangular shaped silicon nitride probe with a microfabricated Pd element at its apex. This thermometer is then connected by two evaporated gold lines and pads. \cite{dobson_new_2007} (K-TEK Nanotechnology) The excitation and demodulation of the AC signals were performed on a HF2LI lock-in amplifier (Zürich instruments). The demodulation of the resulting sideband modulation was done directly using an amplitude/frequency modulation add-on implemented on the lock-in amplifier, assuring the syncing of the oscillators.

\subsection{Diode IV characteristic and thermal response}
\label{sup:diodeIV}
\begin{figure}[h!]
    \centering
    \includegraphics[width=0.65\linewidth]{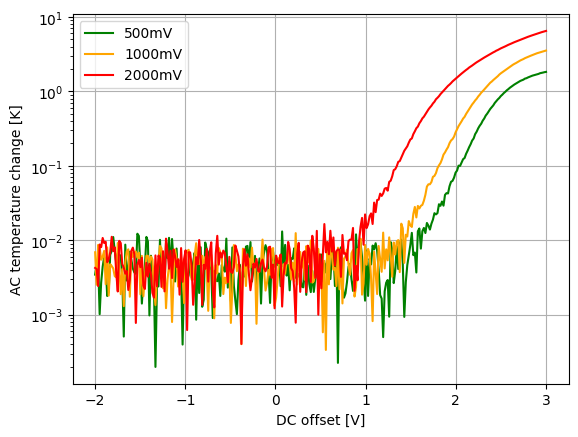}
    \caption{Temperature change on the surface extracted from the first demodulated sideband with respect to the DC offset applied on the diode driving signal. The different colours indicate different driving amplitudes.}
    \label{fig:offset}
\end{figure}

In the case of a diode under a unipolar excitation, only the part of the voltage signal that reaches the switching voltage produces an effective heating. This is demonstrated at Figure \ref{fig:offset} where the sideband demodulated amplitude is plotted against the DC offset of the sine wave for different driving peak-to-peak amplitudes. For a twice greater driving amplitude, the switching point of the diode will be reached for an offset that is lower by half the peak-to-peak amplitude which is verified here. In addition, the curves reproduce the exponential switching of the diode under study followed by a linear increased in the forward passing regime. The switching voltage obtained is around 2.3~\unit{\volt}. This is in-line with IV characteristic of the diode, see Figure~2 and this confirms the importance of considering the non-linearity of the heater in this study.\\

\subsection{Sideband detection of self-heating in a pure resistance driven with unipolar signal}
\label{sup:model}
This section will provide a simplified picture, useful to understand the sideband detection of thermal wave that has been introduced in this work. This section assumes a pure resistive behaviour for the heater to simplify the equations. The proper signal treatment, considering a non-uniform power spectrum arising from the diode, is provided in the main text.

If one assumes that the heater behaves as a simple resistance, the Joule heating generated can be written as :
\begin{equation}
    Q = \frac{V_\mathrm{mod}^2}{2R}
\end{equation}
where Q is the total heat generated on the whole element, V is the voltage signal applied on it and R is its resistance.
For an unipolar sine wave driving signal, at an angular frequency $\omega_m$ :
\begin{equation}
    V_\mathrm{mod}(t) = V_{mod,0} \big(1 + \sin(\omega_1 t + \phi)\big)
\end{equation}
where $V_{mod,0}$ is the amplitude of the signal and $\phi$ an arbitrary phase. For the following development, one assumes that the phase of the probe's reading signal is set to 0 while the other signal are phased compared to it. Thus the temperature variation on the surface generated by the diode can be written as :
\begin{align}
    \Delta T_s(t,x,y) &= \gamma(x,y) \frac{V_{mod,0}^2}{2R}\big[\frac{3}{2} + 2\sin(\omega_1 t + \phi - \omega_1 t_p)\\
    &-\frac{1}{2}\sin(2\omega_1 t + 2\phi - 2\omega_1 t_p)\big]\\
\end{align}
where $\gamma$ is a factor that links the heat generated by the diode below the surface to an actual temperature change at the interface with the probes, it contains the geometrical spreading of the heat, the interfaces etc. $t_p$ is the propagation time of the heat wave from the self-heating diode to the probe, thus the angular frequency of the heat wave multiplied by this time gives the phase lag accumulated by the heat wave. As can be seen, the unipolar excitation of the diode (treated as a resistance $R$) gives rise to a temperature signal in DC, first and second harmonic frequency windows.

Once the wave has reached the probe, the temperature variation on the surface modulates the amplitude of the SThM signal read by the thermometer. This can be understood as an amplitude modulation of the voltage across the Wheatstone bridge which is ultimately the signal measured. The intermixing with the probe's high frequency driving signal arises. This intermixing produces the sidebands that are used for detection.

\begin{equation}
    V_\mathrm{car} = V_{car,0} \sin(\omega_c t)
\end{equation}
\begin{equation}
    V_\mathrm{car}^{AM} =  V_\mathrm{car}\Delta T_s \xi
\end{equation}
The resulting amplitude modulated signal $V_\mathrm{car}^{AM}$ can be written as follows:

\begin{equation}
    \begin{split}
        V_\mathrm{car}^{AM} &= \xi\gamma\frac{V_{mod,0}^2V_{car,0}^{}}{2R}\Big[\sin(\omega_c t)\\
        & + \cos \big((\omega_c-\omega_{1f})t - \phi - \omega_{1f} t_p\big)\\
        & - \cos \big((\omega_c+\omega_{1f})t + \phi + \omega_{1f} t_p\big)\\
        & - \frac{1}{4}\sin\big((\omega_c-\omega_{2f})t-2\phi-\omega_{2f}t_p\big)\\
        & - \frac{1}{4}\sin\big((\omega_c+\omega_{2f})t+2\phi+\omega_{2f}t_p\big)\Big]\\
    \end{split}
    \label{eq:full_signal}
\end{equation}
$\xi$ is the sensitivity of the SThM probe, expressed as the expected change of voltage of the thermometer element for a 1~K temperature change on the surface. $\gamma$ is the temperature change on the surface for one volt applied on the heating diode. $\omega_c$, $\omega_{1f}$ and $\omega_{2f}$ are respectively the angular frequencies of the carrier signal, of the first and second harmonics of the diode's excitation. $V_\mathrm{car}$ and $V_\mathrm{mod}$ are respectively the amplitudes of the voltage used for the reading of the Wheatstone bridge, \textit{i.e.} the carrier signal, and the modulation of its amplitude created by the oscillating temperature on the surface. $\phi$ is an arbitrary phase lag between the oscillator responsible for the probe reading and the oscillator sourcing the diode's excitation voltage. $t_p$ is the propagation time from the heating element to the SThM probe, which depends the distance between them, the material properties as well as its interfaces.

The first harmonic and second harmonic terms each yield their own sub-bands while the DC component gives a component at the carrier frequency, which will appear in the SThM signal at 91\unit{\kilo\hertz}. We can also see that, for an unipolar driving signal, the second harmonic generation is expected to be a fourth of the sub-band signal when considering a perfect resistor. This is qualitatively verified when one sweeps the dc offset on the signal, on Figure \ref{fig:offset}

\subsection{Details about the sample}
\label{sup:sample_details}
The surface regions above the metallic islands (and the pad edges) have an elevated height, approximately $3 \,\unit{\um}$ increased from the inter-island regions, consistent with the $2.8 \,\unit{\um}$ Al top metal thickness noted for chip-package interaction (CPI) qualification of the 22FDX process \cite{klewer2019qual}. Between the top Al and the DTSCR at the FEOL, a metal stack consistent with one of the CPI-qualified BEOL options for 22FDX \cite{klewer2019qual} is used: 7 layers of thin metal with low-K/ultra-low-K (ULK) dielectric, 1 layer of thick metal with TEOS (tetraethyl orthosilicate) dielectric, and 1 layer of ultra-thick metal (UTM) with TEOS dielectric. Specification of TEOS typically corresponds to its use in manufacturing as a precursor for an oxide such as SiO$_2$ which would then act as the inter-metal dielectric for those layers \cite{bulla1998teos}. Each of these buried metal layers is Cu, and the height of the top of the UTM layer relative to the active device layer (e.g. the DTSCR) is $\sim6.9 \,\unit{\um}$ as given for 22FDX metal stack \#11 \cite{gf2020stackup}. Each of these metal layers has a similar fill of smaller floating metallic islands along with some signal routing lines, but the presence of inter-metal dielectric means that these metal patterns do not have a significant effect on the chip's physical surface topography. Beneath the diode, the backplane silicon substrate of the chip is $254 \,\unit{\um}$ thick and mounted to a PCB using a silver adhesive.

The DTSCR power dissipation is distributed non-uniformly over a significant area. The junction area of the three diodes in Figure~2a are given as $A_{D1}=21.12\,\unit{\um}^2$ and $A_{D2}=A_{D3}=3.52\,\unit{\um}^2$. While the junction area of diode D1 is greater than the other two, similar power levels are dissipated in each of the three diodes due to the same current passing through all three diodes and only a minor difference in forward-bias voltage resulting from the larger area of D1. These three junction areas are each distributed across multiple parallel "finger" devices laid out across a $\sim250\,\unit{\um}^2$ region denoted via the blue rectangle in the micrograph presented in Figure~3a.

\newpage
\subsection{Amplitude-dependent sideband-SThM line scans}
\label{sup:amp_dependent}
\begin{figure}[h!]
    \centering
    \includegraphics[width=0.92\linewidth]{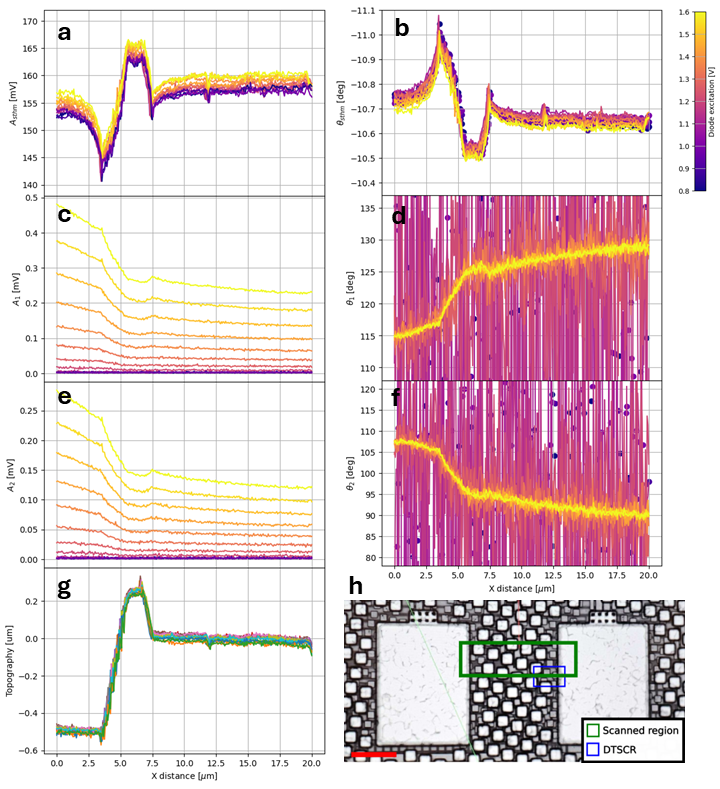}
    \caption{Power-dependent study on the buried diode. The color scale presented on the top right denotes the amplitude of the voltage excitation applied on the diode. \textbf{a} and \textbf{b} are respectively the amplitude and phase of the signal at $91$\si{\kilo\hertz}. \textbf{c} and \textbf{d} are respectively the amplitude and phase traces of the first harmonic sideband at $92$\si{\kilo\hertz}. \textbf{e} and \textbf{f} are respectively the amplitude and phase traces of the second harmonic sideband at $93$\si{\kilo\hertz}. \textbf{g} is topography traces obtained for the different powers. \textbf{h} presented the micrograph of the region of interest, the dashed line indicates where the traces were recorded. The shaded red region is the place where the heating diode is buried as identified by the maps presented at Figure 2 of the main article.}
    \label{fig:sup_power_dependent}
\end{figure}

At Figure S\ref{fig:sup_power_dependent}, different SThM traces have been recorded while increasing the diode's driving voltage amplitude. As can be seen, the amplitude of both the first and second harmonic sideband are progressively increased while the phase does not depend on this amplitude. As the phase signal reflects the propagating time of the thermal waves, this was expected and acts as the confirmation of the relevance of our thermal wave analysis, provided in the main text.

\newpage
\bibliographystyle{IEEEtran}
\bibliography{mybib}